\documentclass[12pt,a4paper,twoside,showpacs,showkeys,nofootinbib]{revtex4}
\usepackage{amsmath,amsfonts}
\usepackage{amsbsy}
\usepackage{mathrsfs}
\usepackage{color}
\usepackage{psfrag}
\usepackage{enumerate}
\usepackage{amsmath,amssymb,calc,amsfonts}
\usepackage{latexsym}
\def\ut#1{\rlap{\lower1ex\hbox{$\sim$}}#1{}}
\DeclareFontFamily{U}{rsfs}{}         
\DeclareFontShape{U}{rsfs}{m}{n}{<5> rsfs5 <6><7> rsfs7          %
  <8><9><10><10.95><12><14.4><17.28><20.74><24.88> rsfs10}{}     %
\DeclareMathAlphabet{\mathfs}{U}{rsfs}{m}{n}                     %
\def\beq{\begin{eqnarray}}
\def\eeq{\end{eqnarray}}

\def\del{\partial}

\begin{document}

\title{Black holes in Einstein-Gauss-Bonnet gravity: dynamical and 4-dimensional novel stationary black hole}
\date{\today}
\author{Bhramar Chatterjee}\email{bhramar.chatterjee@gmail.com}

\begin{abstract}
We consider black hole in Einstein-Gauss-Bonnet(EGB) gravity in a dynamical setup and also the stationary black hole in recently discovered\cite{Glavan} $4$-dimensional novel Einstein-Gauss-Bonnet gravity. For dynamical EGB black hole in completely local settings we calculate the temperature associated with the horizon. For the stationary black hole in $4$-dimensional novel EGB theory we explore the thermodynamic properties including calculation of temperature and entropy.
\end{abstract}



\maketitle
\section{Introduction}
Alternative theories of gravity have been a prevalent subject of current research in gravitational physics. The motivation behind this is to look for answers to some fundamental problems which can not be resolved in Einstein's general relativity (GR), for example the singularity. Since inclusion of higher order curvature terms in the action as corrections seems natural progression from Einstein's GR, many such theories exist, and Einstein-Gauss-Bonnet gravity(EGB) which comprises of the Gauss-Bonnet term (quadratic in curvature) added to the Lagrangian of Einstein-Hilbert action is one of the most promising of such theories \cite{Boulware, Zumino}. This particular combination of quadratic curvature terms appears as the second order correction in the action of Lanzcos-Lovelock gravity\cite{Lovelock, Lovelock1, Lanzcos},the zeroth and first order terms being the cosmological constant and Einstein-Hilbert action respectively. The advantage of Lanzcos-Lovelock gravity is that the equations of motion contain only up to second derivative of metric even if higher curvature terms are present in the action, which makes the theory ghost free. Also, Gauss-Bonnet term appears in the low energy limit of heterotic string theory\cite{Zwiebach,Gross,Gross1,Metsaev,Metsaev1} which is another reason to consider this theory as an alternative to Einstein's general relativity.\\
Now, the effect of these higher order curvature terms is expected to become apparent in extremely curved domain of spacetime. So study of black holes in Einstein-Gauss-Bonnet gravity can reveal different features from Einstein's gravity. Recent availability of high precision observational data in the strong gravity regime around black holes made it possible to study different phenomena related to black holes more thoroughly and precisely\cite{Berti,Barack}. So this is a good time to test alternative theories of gravity in the background of black holes compared to standard general relativity.\\
Stationary black holes has been studied exhaustively over the years, particularly for their thermodynamic properties. But black holes in reality rarely reach equilibrium. A physical black hole may grow by absorbing nearby matter or coalesce with another black hole or emit Hawking radiation, all of which are highly dynamical processes. Moreover the notion of black hole event horizon, being a global one, is not at all suitable for practical purposes. To locate the event horizon of a black hole one has to know the entire structure of the spacetime beforehand, which is not possible always. For all of these reasons some local definitions of black hole horizons which incorporate the dynamical nature of the system has been proposed over the last few years with varying degrees of success. Examples include trapping horizon\cite{Hayward1, Hayward2}, slowly evolving horizon\cite{Booth}, dynamical horizon\cite{Ashtekar1, Krishnan} and isolated horizon\cite{Ashtekar2}. All these horizons are locally defined, i.e unlike event horizon only local properties of the spacetime are sufficient to locate the horizon. Also these horizons are endowed with nice properties i.e they obey the laws of black hole mechanics and a temperature and entropy can be associated with them.\\

In the first part of this work we will consider a dynamical black hole in Einstein-Gauss-Bonnet gravity in the context of Hayward's trapping horizon. For a $4$-dimensional spacetime with spherical symmetry the trapping horizon is defined in a $2 + 2$ framework using only the local geometry without any assumptions on the global developments of the spacetime in which the horizon is embedded. For Einstein-Gauss-Bonnet gravity, trapping horizon and it's properties has been studied in detail in \cite{Nozawa}. There are two branches of solutions in EGB gravity : the GR branch which has an asymptotic general relativistic limit, and the non-GR branch which has no such limit. It is found that for the GR branch, for a $n$-dimensional spacetime which has symmetries corresponding to a $(n - 2)$-dimensional maximally symmetric space with constant curvature, the trapping horizon has very similar behavior to that of in general relativity under the null energy condition, whereas the non-GR branch solution behaves completely differently. Since the spacetime is dynamical, there is no timelike Killing vector, but because of the maximally symmetric nature of the spacetime one can define a Kodama vector \cite{Kodama}, in a similar manner to that of spherically symmetric $4$-dimensional dynamical spacetime, which plays an equivalent role of Killing vector. We shall calculate the temperature of the trapping horizon in EGB gravity by considering radiation from the horizon in the form of positive frequency field modes of the Kodama vector. This method has previously been employed to calculate temperature of horizons in different $4$-dimensional spacetimes \cite{Chatterjee1,Chatterjee2,Chatterjee3}. The set up is essentially local and dynamical. The nature of the trapping horizon is characterized by the energy conditions, without specifying the energy momentum tensor of the matter field.\\

For the second part of this paper we will consider the stationary black hole solution in the recently discovered $4$-dimensional novel Einstein-Gauss-Bonnet gravity \cite{Glavan}. Following Lovelock's theorem \cite{Lovelock, Lovelock1}, Einstein's GR with the cosmological constant is the unique theory of gravity in $4$-dimension if we assume diffeomorphism invariance, metricity and second order equations of motion. The Gauss-Bonnet term, which is the next higher order term in curvature, becomes a total derivative in $4$-dimensions, hence does not appear in the Einstein's equations. However, recently Glavan and Lin \cite{Glavan} has proposed a modified theory of gravity which has Gauss-Bonnet term contributing in the $4$-dimensional gravitational dynamics. According to \cite{Glavan}, this theory is formulated in $D > 4$ dimensions and the action consists of Einstein-Hilbert term plus the cosmological constant and the Gauss-Bonnet term multiplied by a factor of $1/(D - 4)$. The $4$-dimensional theory is recovered as the limit $D \rightarrow 4$ at the level of equations of motion and there is significant contribution from the Gauss-Bonnet term. The choice of action is ad hoc at the best but this theory has some very interesting features. Particularly the vacuum black hole solution which behaves like the Schwarzschild black hole at large distances is very different from Schwarzschild solution near the horizon. This black hole solution admits two horizons, outer and inner, despite being static and without any charges and the gravitational potential inside the inner horizon becomes repulsive,  which suggests a white hole spacetime inside the black hole. There has been a lot of activities concerning the black hole in this $4$-dimensional novel Einstein-Gauss-Bonnet gravity \cite{Fernandes, Ai, Mukhoyama, Konoplya, Ghosh, Ghosh1, Dadhich}. In this paper we shall construct double null coordinates for this black hole and calculate the temperature, entropy and other thermodynamic properties associated with the black hole horizon.  \\

The plan of the paper is as follows : in the first section we will discuss basics of Einstein-Gauss-Bonnet gravity briefly starting from the action. The dynamical black hole will be discussed in the next section. After introducing the concept of future outer trapping horizon (FOTH, which characterizes a black hole in dynamical scenario) for higher dimensional spacetime and Kodama vector, we shall construct the field modes of Kodama vector and calculate the probability current which will give the temperature of the horizon. In the last section we will discuss $4$-dimensional novel EGB black hole and it's thermodynamic properties and conclude. We have chosen $c = \hbar = 1$.

\section{Einstein Gauss-Bonnet gravity}
We begin with a brief discussion of Einstein-Gauss-Bonnet(EGB) gravity. In the presence of a cosmological constant the action is given by,
\beq
S = \int d^{n}x \sqrt{-g}\Big[\frac{1}{16\pi G_n}(R - 2\Lambda + \alpha L_{GB})\Big] + S_{matter},\label{action}
\eeq
where $G_n$,$R$ and $\Lambda$ are the $n$-dimensional gravitational constant, Ricci scalar and the cosmological constant respectively. The Gauss-Bonnet term $L_{GB}$ ,which is quadratic in curvature, is a combination of Ricci scalar, Ricci tensor $R_{\mu\nu}$ and Riemann tensor $R_{\mu\nu\alpha\beta}$ as
 \beq
L_{GB} = R^2 - 4 R_{\mu\nu}R^{\mu\nu} + R_{\mu\nu\alpha\beta}R^{\mu\nu\alpha\beta}.
 \eeq
 $\alpha(\geq 0)$ is the Gauss-Bonnet coupling constant and $S_{matter}$ is the contribution from matter field. In $4$-dimensions $L_{GB}$ does not contribute to the field equations since it becomes a total derivative. For $D > 4$, the field equation derived from the action (\ref{action}) is
 \beq
 G^\mu_\nu + \alpha H^\mu_\nu + \Lambda \delta^\mu_\nu = 8\pi G_n T^\mu_\nu,\label{feq}
 \eeq where,
 \beq
 G_{\mu\nu}= R_{\mu\nu} - \frac{1}{2}g_{\mu\nu}R,
 \eeq
 \beq
 H_{\mu\nu} = 2[RR_{\mu\nu}- 2R_{\mu\alpha}R^{\alpha}_{\nu} - 2 R^{\alpha\beta}R_{\mu\alpha\nu\beta}+ R_{\mu}^{\alpha\beta\gamma}R_{\nu\alpha\beta\gamma}] - \frac{1}{2}g_{\mu\nu}L_{GB}.
 \eeq
  The field equations (\ref{feq}) contain only up to linear second derivative of the metric. $T_{\mu\nu}$ is the energy-momentum tensor of the matter fields. Now following \cite{Nozawa}, if we assume the $n$-dimensional spacetime $(M^n, g_{\mu\nu})$ is a warped product of a $(n-2)$-dimensional constant curvature space $(K^{n-2}, \gamma_{ij})$ and a $2$-dimensional orbit spacetime $(M^2, g_{ab})$ under the isometry of $(K^{n-2}, \gamma_{ij})$ , the metric can be written as,
  \beq
g_{\mu\nu}dx^{\mu}dx^{\nu} = g_{ab}(y)dy^{a}dy^{b} + r^2(y)\gamma_{ij}dz^{i}dz^{j},
\eeq
where $a, b = 0,1; i, j = 2, ..., n-1$, $\gamma_{ij}$ is the unit metric on $(K^{n-2}, \gamma_{ij})$ with sectional curvature $k = -1, 0, +1$, $r$ is a scalar on $(M^2, g_{ab})$ with the boundary at $r = 0$. If there are no matter fields, we will get the vacuum solution which is given by the generalized Boulware-Deser-Wheeler solution\cite{Boulware, Wheeler, Cai, Soh}
  \beq
  ds^2 = -F(r)dt^2 + \frac{dr^2}{F(r)} + r^2 \gamma_{ij}dz^{i}dz^{j},
  \eeq
   with
    \beq F(r) = k + \frac{r^2}{2\tilde\alpha}\Big[1 \mp \sqrt{1 + \frac{64\pi G_n \tilde\alpha m}{(n-2)V^k_{n-2}r^{n-1}}+ 4\tilde\alpha\tilde\Lambda}\Big] ,
    \eeq where $\tilde\alpha = (n-3)(n-4)\alpha, \tilde\Lambda = \frac{2\Lambda}{(n-1)(n-2)}$. $V^k_{n-2}$ is the area of the unit $(n-2)$-dimensional space with constant curvature and $m$ is the gravitational mass of the solution. The upper sign in $F(r)$ refers to the GR branch because in the asymptotic limit this gives a Schwarzschild-like solution. The lower sign is the non-GR branch.\\
    When matter fields are present, the setup becomes dynamic. There are no longer an event horizon or a Killing vector. But if the spacetime is maximally symmetric, as we have assumed here, one can define a trapping horizon which characterizes a black hole using only local geometry. This we will discuss in the next section.
\section{Dynamical EGB black hole}
To describe the trapping horizon of a dynamical black hole in EGB gravity it is convenient to write the metric in double null coordinates as,
\beq
ds^2 = -2 e^{-f(u,v)} dudv + r^2(u,v)\gamma_{ij}dz^i dz^j.\label{dynmet}
\eeq
The null vectors $\del/\del u$ and $\del/\del v$ are taken to be future directed with $\del/\del u (\del/\del v)$ in the inward(outward) direction. The expansions of the two radial null geodesics are given by,
\beq
\theta_+ = (n-2)\frac{\del_v r}{r},\\
\theta_- = (n-2)\frac{\del_u r}{r}
\eeq
Now a marginal surface in this spacetime is a $(n-2)$-surface with $\theta_+\theta_- = 0$ and without any loss of generality we can choose $\theta_+ = 0$. A future outer trapping horizon (FOTH) which corresponds to a black hole horizon in this spacetime is a $(n-1)$- hypersurface foliated by marginal surfaces such that on each $(n-2)$-marginal surfaces $\theta_- < 0$ and $\del_u \theta_+ < 0$. This implies the ingoing null rays are converging $(\theta_- < 0)$, the outgoing null rays are converging inside the FOTH $(\del_u \theta_+ < 0)$, instantaneously parallel to the horizon $(\theta_+ = 0)$ and diverging just outside the horizon , which indicates a black hole spacetime.\\
 The active gravitational mass of the trapping horizon is the generalized Misner-Sharp mass $m$. Since the equations of motion contain quadratic curvature terms, $m$ is quadratic in $e^{f}\theta_+\theta_-$ and is given by \cite{Nozawa},
 \beq
 \frac{2}{(n-2)^2}r^2e^{f}\theta_+\theta_- = -k -\frac{r^2}{2\tilde\alpha}\Big[1 \mp \sqrt{1 + \frac{64\pi G_{n}\tilde\alpha m}{(n-2)V^k_{n-2}r^{n-1}} + 4\tilde\alpha\tilde\Lambda}\Big]
 \eeq
 The "$-$" sign in front of the square root corresponds to the GR branch solution which has a Schwarzschild like limit as $\alpha \rightarrow 0$. The "$+$" sign gives the non-GR branch which behaves rather pathologically. The FOTH in the GR branch has very similar properties as that of in general relativity under the null energy condition (for details see \cite{Nozawa}) and in the next subsection we will calculate temperature of this horizon.
 \subsection{Kodama vector and surface gravity}
 For a dynamical spacetime there is no Killing vector, so definition of time becomes ambiguous. But in $4$-dimension if the spacetime is spherically symmetric, Kodama vector $(K)$ generates a preferred flow of time \cite{Kodama}. Using Einstein's equations it can be shown that the vector field $ J^a = G^{ab}K_b$ is divergence free where $G^{ab}$ is the Einstein tensor \cite{Kodama,Visser}. This implies that even though the spacetime is completely dynamical, $J^a$ is the locally conserved energy flux. So the Kodama vector provides a preferred timelike direction. Also, it is shown in \cite{Hayward4}, that the Noether charge associated with the Kodama current $J^a$ leads to the Misner-Sharp mass \cite{Misner} of the spacetime. For $(n \geq 5)$-dimension, if the spacetime is maximally symmetric as we have assumed here, we can define a generalized Kodama vector by,
 \beq
 K^a = \epsilon^{ab}D_{b}r,
 \eeq
 where $\epsilon^{ab}$ is a volume element of $(M^2 , g_{ab})$ and $D_a$ is a linear connection compatible with the metric in $(M^2 , g_{ab})$. This vector lies entirely on the transverse $(u,v)$ plane and using the metric (\ref{dynmet}) can be written as,
 \beq
 K = e^f(\del_v r \del_u - \del_u r \del_v).\label{Kodama}
 \eeq
 The dynamical surface gravity of the trapping horizon $(\kappa_{TH})$ is defined as,
 \beq
 K^b D_{[b}K_{a]} = \kappa_{TH}K_a.
 \eeq
 So, using (\ref{dynmet}) we have,
 \beq
 \kappa_{TH} = \frac{1}{2}D^2 r = -e^f \del_u \del_v r.
 \eeq
The FOTH condition $(\del_u \theta_+ < 0)$ makes sure that $\kappa_{TH}$ is positive on the FOTH. We will show that $\kappa_{TH}$ is indeed the temperature of the FOTH using field modes of Kodama vector $(K)$ in the following subsection.
 \subsection{Field modes and probability current}
 From the expression of Kodama vector (\ref{Kodama}), it is easy to see that any smooth function of $r$ is a zero mode of $K$. Now we can evaluate other positive frequency eigenmodes of $K$ using,
 \beq
 i K Z_\omega = \omega Z_\omega.
 \eeq
 Here, $Z_\omega$ are the eigenfunctions corresponding to the positive frequency $\omega$. The method of finding eigenmodes of Kodama vector has been discussed in detail in \cite{Chatterjee2, Chatterjee3}. Following these one can find
\begin{equation}\label{umode}
 Z_\omega= C(r)\begin{cases} \theta_+^{-\frac{i\omega}{\kappa_{TH}}} &
\text{for}\;\theta_+>0\\
(|\theta_+|)^{-\frac{i\omega}{\kappa_{TH}}} & {\rm for}\;
\theta_+<0.\end{cases}
\end{equation}
where the $(n-2)$-surfaces are not trapped `outside the trapping horizon' ($\theta_+>0$) and fully trapped `inside' ($\theta_+<0$). The only assumption made in this calculation is that $\kappa_{TH}$ is a slowly varying function in some small neighbourhood of the FOTH. These are precisely the modes which
are defined outside and inside the dynamical horizon respectively but not on the horizon. Now we have to keep in mind that the modes (\ref{umode}) are not
ordinary functions, but are distribution-valued. And as distributions (or generalized functions) these are perfectly well behaved on the horizon. Using the standard results \cite{Gel'fand},the appropriate distributions for the $Z_\omega$ are,
\beq Z_\omega = \lim_{\epsilon \to 0} C(r)(\theta_{+} + i\epsilon)^{-\frac{i\omega}{\kappa_{TH}}}
= \left\{ \begin{array}{ll}
           C(r)\left(\theta_{+}\right)^{-\frac{i\omega}{\kappa_{TH}}} & \mbox{ for $\theta_{+} > 0$,} \\  C(r)|\theta_{+}|^{-\frac{i\omega}{\kappa_{TH}}}
e^{\frac{\pi\omega}{\kappa_{TH}}} & \mbox{ for $\theta_{+} < 0$. }
            \end{array}
\right.\label{plusdist}
\eeq  The
distribution (\ref{plusdist}) is well-defined for all values of $\theta_+$ and $\kappa_{TH}$, and it is differentiable to all orders. The modes $Z^*_\omega$ are
given by the complex conjugate distribution
\beq Z_{\omega}^* = \lim_{\epsilon \to 0} {C(r)}^*(\theta_{+} - i\epsilon)^{\frac{i\omega}{\kappa_{TH}}}
= \left\{ \begin{array}{ll}
           {C(r)}^*\left(\theta_{+}\right)^{\frac{i\omega}{\kappa_{TH}}} & \mbox{ for $\theta_{+} > 0$,} \\  {C(r)}^*|\theta_{-}|^{\frac{i\omega}{\kappa_{TH}}}
e^{\frac{\pi\omega}{\kappa_{TH}}} & \mbox{ for $\theta_{+} < 0$. }
            \end{array}
\right.\label{minusdist}
\eeq
Let us now calculate the probability density in a single particle Hilbert space for positive frequency solutions across the dynamical horizon along the direction of the Kodama vector $K$,
\begin{equation}\label{numberd}
 \varrho(\omega)= -\frac{i}{2}\Big[Z^*_\omega KZ_\omega-K Z^*_\omega Z_\omega\Big]=\omega Z_\omega^*Z_\omega.
\end{equation}
A straightforward calculation gives, apart from a positive function of $r$,
\begin{eqnarray}
\varrho(\omega) &=& \omega(\theta_++i\epsilon)^{-\frac{i\omega}{\kappa_{TH}}} (\theta_+-i\epsilon)^ {\frac{i\omega}{\kappa_{TH}}}.\\ \nonumber
 &=& \begin{cases} \omega & \text{for}\;\theta_+>0\\
\omega e^{\frac{2\pi\omega}{\kappa_{TH}}} & \text{for}\;\theta_+<0.\end{cases}
\end{eqnarray}
The conditional probability that a particle emits when it is incident on the horizon from inside is,
\beq
 P_{(emission|incident)} = \frac{P_{(emission{\cap}incident)}}{P_{(incident)}} =  e^{-\frac{2\pi\omega}{\kappa_{TH}}}
\eeq
comparing this with the Boltzmann factor $e^{-\beta\omega}$ gives the temperature of this dynamical horizon  $ T_H = \kappa_{TH}/2\pi$, which is the desired result. This temperature for the Gauss-Bonnet trapping horizon is obtained for the GR branch solution only, which is very similar to the temperature of trapping horizon in $4$-dimension \cite{Chatterjee2}. The effect of the Gauss-Bonnet term is embedded in $\kappa_{TH}$ which is indeed different from it's $4$-dimensional counterpart since the spacetime metrics are different. But otherwise the trapping horizon for the GR branch in EGB gravity behaves very similarly to the trapping horizon in $4$-dimensional GR. For the non-GR branch the trapping horizon behaves quite differently under the null energy condition, and the results are not easy to interpret.
\section{The novel 4D-EGB black hole}
According to \cite{Glavan}, the $4$-dimensional novel EGB theory is formulated in $n > 4$ dimensions with the action ,
\beq
S = \int d^{n}x \sqrt{-g}\Big[\frac{1}{16\pi G_n}(R - 2\Lambda + \frac{\tilde\alpha}{n-4} L_{GB})\Big] ,\label{faction}
\eeq
and the $4$-D theory is recovered in the limit $n \rightarrow 4$ at the level of equations of motion. For a static spherically symmetric $4$-D spacetime with a vanishing cosmological constant this leads to the solution,
\beq
ds^2 = -f(r)dt^2 + \frac{dr^2}{f(r)} + r^2 d\Omega^2,\label{mtrc}
\eeq
where
\beq
f(r) = 1 + \frac{r^2}{2\alpha}\Big(1 \mp \sqrt{1 + \frac{8\alpha M}{r^3}}\Big).
\eeq
Here $d\Omega^2$ is the usual metric on unit 2-sphere, $\alpha$ is rescaled GB coupling constant $(\alpha = 16\pi\tilde\alpha)$ and $M$ is mass of isolated body. We have also chosen $G = 1$. In the asymptotic limit $(r\rightarrow\infty)$ the $"+"$ sign in $f(r)$ gives Schwarzschild-de Sitter black hole with negative gravitational mass and the $"-"$ sign gives  Schwarzschild black hole with positive gravitational mass. So we choose $"-"$ sign in $f(r)$ for attractive gravity. This $f(r)$ has roots at
\beq
r_{\pm} = M \pm \sqrt{M^2 - \alpha}.
\eeq
For $\alpha > 0$, the roots are real for $M^2 > \alpha$, while $M^2 = \alpha$ gives the extremal case. So unlike the Schwarzschild solution, in this case the black hole has two horizons : outer$(r_+)$ and inner$(r_-)$, just like the Reissner-Nordstr\"{o}m case where the electric charge $Q^2$ is replaced by GB coupling constant $\alpha$ which is quite exceptional for a static uncharged black hole. As was shown in \cite{Glavan}, the gravitational potential has a minimum and is attractive to the right and repulsive to the left of this minimum. For $M^2 > \alpha$, where two horizon exist this means the outer horizon is the event horizon of a black hole while the inner horizon envelopes a white hole spacetime inside. In the following subsections we will explore the thermodynamic properties of the black hole particularly the temperature and entropy associated with the outer horizon. To calculate temperature we will use the formalism adopted for calculating temperature of dynamical EGB black hole in the previous section.For this we need the metric in the double null Kruskal coordinates so that the coordinate singularity can be removed.
\subsection{Double-null coordinates}
   Because of the presence of coordinate singularities at $r_\pm$, Kruskal coordinates are required to extend the metric across these surfaces. However one fact has to be mentioned clearly i.e a single Kruskal patch will not cover the entire manifold in this case due to the presence of two horizons. The Kruskal coordinates we construct to remove the singularity at the outer horizon $(r_+)$ fails to be regular at the inner horizon $(r_-)$ and we need a different set of coordinate transformation for the inner horizon. This is similar to  Reissner-Nordstr\"{o}m black hole again. Let us construct the Kruskal coordinates across the outer horizon (the black hole event horizon).
Near $r = r_+$ \cite{Poisson},
\beq
f(r) \approx f'(r)\mid_{r_+}(r-r_+) = 2\kappa_{+}(r-r_+)
\eeq
where $\kappa_+ \equiv \frac{1}{2}f'(r_+).$ Then we have,
\beq
r_* = \int \frac{dr}{f(r)} = \int \frac{dr}{2\kappa_+ (r-r_+)} = \frac{1}{2\kappa_+}\ln |\kappa_+(r-r_+)|.
\eeq
Now we can define null coordinates $u = t - r_*$, and $v = t + r_*$, and the Kruskal coordinates are defined as $U_+ = \mp e^{-\kappa_+ u}$ and $V_+ = e^{\kappa_+v}$. The upper sign in $U_+$ is for $r > r_+$, and the lower sign is for $r < r_+$. In these coordinates $f(r) \simeq -2U_+ V_+$ near $r_+$ and the metric becomes,
\beq
ds^2 = \frac{f(r)}{\kappa_+^2 U_+ V_+}d U_+ d V_+ + r^2d\Omega^2 \simeq - \frac{2}{\kappa_+^2}d U_+ d V_+ + r^2 d\Omega^2.
\eeq
The future outer horizon is defined as $U_+ = 0, V_+ > 0$. This metric is regular at the outer horizon. After crossing the outer horizon as one approaches the inner horizon $r_-$, $r_* \rightarrow \infty$, which implies $v-u \rightarrow \infty, $ i.e $U_+ V_+ \rightarrow \infty.$ So these Kruskal coordinates are singular at inner horizon and should be used only in an interval $r_1 < r < \infty$ where $r_1 > r_-$ is some cutoff radius.
\subsection{Surface gravity}
Since the spacetime is static, $K = \del_t$ is a Killing vector and the event horizon is also a Killing horizon. Surface gravity $(\kappa)$ is defined using the Killing vector $(K^\mu)$,
\beq
\left(-K^\mu K_\mu\right);_\nu = 2\kappa K_\nu.
\eeq
And in this case of 4D EGB black hole the surface gravity associated with the outer horizon is found to be $\kappa_+ = \frac{1}{2}f'(r_+)$. It also satisfies the extremal limit i.e $\kappa_+ = 0 $ when $\alpha = M^2$.
\subsection{Horizon temperature}
To determine the temperature of the outer horizon we shall adopt the formalism discussed in \cite{Chatterjee1} for static spherically symmetric black holes. Considering a scalar field, we shall construct the modes of the Killing vector across the horizon. The positive frequency modes$(\omega > 0)$ of the Killing vector satisfy,
\beq
i\del_t \Phi_\omega = \omega\Phi_\omega.\label{KE}
\eeq
In Kruskal coordinates $\del_t$ becomes,
\beq
\del_t = -\kappa_+ U_+ \del_{U_+} + \kappa_+ V_+ \del_{V_+},
\eeq
and the modes are separated,
\beq
\Phi_\omega = f_\omega (U_+) + g_\omega (V_+).
\eeq
Equation (\ref{KE}) can be solved immediately, and the solutions are,
\beq
f_\omega (U_+) = N_\omega |U_+|^\frac{i\omega}{\kappa_+}\\
g_\omega (V_+) = N_\omega \left(V_+\right)^{-\frac{i\omega}{\kappa_+}}.
\eeq
The ingoing $V_+$ modes are well behaved and of no interest to us. The outgoing $U_+$ modes have a logarithmic singularity at the horizon and we need these modes only to calculate the outgoing probability current.
To remove the singularity at the horizon we have used the fact that the field modes are essentially distributions, and the distributions are well behaved at the horizon. The appropriate form of the distribution is\cite{Gel'fand},
\beq
 f_{\omega} = \lim_{\epsilon \to 0}N_{\omega}|U_{+} + i\epsilon|^{\frac{i\omega}{\kappa_{+}}}
= \left\{ \begin{array}{ll}
           N_{\omega}\left(U_{+}\right)^{\frac{i\omega}{\kappa_{+}}} & \mbox{ for $U_{+} > 0$,} \\  N_{\omega}|U_{+}|^{\frac{i\omega}{\kappa_{+}}}
e^{-\frac{\pi\omega}{\kappa_{+}}} & \mbox{ for $U_{+} < 0$, }
            \end{array}
\right.
\eeq
and the complex conjugate form is
\beq
 \overline{f_{\omega}} = \lim_{\epsilon \to 0}N_{\omega}^*\,|U_{+} - i\epsilon|^{-\frac{i\omega}{\kappa_{+}}}
= \left\{ \begin{array}{ll}
       N_{\omega}^*\left(U_{+}\right)^{-\frac{i\omega}{\kappa_{+}}} & \mbox{ for $U_{+} > 0$,} \\ N_{\omega}^*\,|U_{+}|^{-\frac{i\omega}
{\kappa_{+}}}
e^{-\frac{\pi\omega}{\kappa_{+}}} & \mbox{ for $U_{+} < 0$. }
            \end{array}
\right.
\eeq
Now, the emission probability is given by the outgoing probability current,
\beq
 j^{out} = -i\left[-\kappa_{+}U_{+}\left(\partial_{U_{+}}\overline{\Phi_{\omega}}\right)\Phi_{\omega}
 + \kappa_{+}U_{+}\overline{\Phi_{\omega}}\left(\partial_{U_{+}}\Phi_{\omega}\right)\right].
\eeq
A straightforward calculation gives,
\beq
 j^{out} = \omega U_+ |N_\omega|^2\lim_{\epsilon \to 0}\left[\frac{1}{U_+ -i\epsilon}+\frac{1}{U_+ + i\epsilon}\right]\left(U_+
-i\epsilon\right)^{-\frac{i\omega}{\kappa_+}}\left(U_+ + i\epsilon\right)^{\frac{i\omega}{\kappa_+}},
\eeq
so that,
\beq
j^{out} = \left\{ \begin{array}{ll}
           |N_{\omega}|^{2} & \mbox{ for $U_{+} > 0$,} \\  |N_{\omega}|^{2}e^{-\frac{2\pi\omega}{\kappa_{+}}} & \mbox{ for $U_{+} < 0$. }
            \end{array}
\right.
\eeq
Then we get the conditional probability that a particle emits from the horizon when it is incident on the horizon from the inside,
\beq
P_{(emission|incident)} = \frac{P_{(emission{\cap}incident)}}{P_{(incident)}} = \frac{|N_{\omega}|^{2}e^{-\frac{2\pi\omega}{\kappa_{+}}}}{|N_{\omega}|^{2}}.
\eeq
Comparing this with the Boltzmann factor, $e^{-\beta\omega}$, we found the temperature of the outer horizon is $\kappa_+/2\pi$. This temperature is not similar to the temperature of a Schwarzschild black hole, rather it looks like the temperature of the outer horizon in Reissner-Nordstr\"{o}m black hole with the Gauss-Bonnet coupling constant playing the role of electric charge. So due to the presence of the Gauss-Bonnet term in $4$-dimension the black hole acquires two horizons. We will discuss this in detail in the next section.
\subsection{Entropy}
In General Relativity black hole entropy $(S)$ is given by the Hawking-Bekenstein formula $S = A/4$, where $A$ is the area of the event horizon. However one cannot naively apply this formula to higher order curvature theories of gravity to obtain entropy. In that case it is most logical to calculate entropy from the first law of black hole thermodynamics $dM = T dS$ which gives\cite{Cai}
\beq
S = \int \frac{dM}{T(M)},
\eeq
So, we get
\beq
S = \int\frac{1}{T(r_+)}\frac{\del M}{\del r_+}dr_+ = S_0 + 2\pi\int_{0}^{r_+} (r_+ + \frac{2\alpha}{r_+})dr_+,
\eeq
where $S_0$ is some integration constant which can be fixed by arguing entropy should be zero when the horizon of the black hole shrinks to zero. Then we find,
\beq
S = \pi r_+^2 + 2\pi\alpha \ln{r_+^2} + \tilde S_0 = \frac{A}{4} + 2\pi\alpha \ln{\frac{A}{A_0}},
\eeq
where $A$ is the area of outer horizon and $A_0$ is some constant with dimension of area. This result is very surprising because though the first term gives standard Hawking-Bekenstein entropy\cite{Bekenstein, Hawking}, nowhere in general relativity the logarithmic term appears in the expression for entropy. It only appears as the leading order correction to entropy in some quantum theories of gravity\cite{Majumdar, Cai1, Cai2, Cognola}.
\subsection{Heat capacity}
The heat capacity is given by,
\beq
C = (\frac{\del M}{\del T}) = (\frac{\del M}{\del r_+})(\frac{\del r_+}{\del T}).
\eeq
In this case we have,
\beq
\frac{\del M}{\del r_+} =  1 - \frac{r_+^2 + \alpha}{2r_+^2} = \frac{r_+^2 - \alpha}{2r_+^2},
\eeq and
\beq
\frac{\del T}{\del r_+} = \frac{1}{4\pi}\frac{2\alpha^2 + 5\alpha r_+^2 - r_+^4}{r_+^2(r_+^2 + 2\alpha)^2},
\eeq which gives,
\beq
C = \frac{2\pi(r_+^2 - \alpha)(r_+^2 + 2\alpha)^2}{2\alpha^2 + 5\alpha r_+^2 - r_+^4}.
\eeq
\subsection{Free energy}
The free energy of black holes is defined as $F = M - TS$, and in this case is given by,
\beq
F = \frac{r_+^2 + \alpha}{2r_+} - \frac{r_+(r_+^2 - \alpha)}{4(r_+^2 + 2\alpha)} - \frac{\alpha(r_+^2 - \alpha)}{2r_+(r_+^2 + 2\alpha)}\ln\frac{A}{A_0}.
\eeq
\section{Discussions}
In this paper we have discussed two different types of black holes in the context of Einstein-Gauss-Bonnet gravity. For both of these black holes, we have concentrated solely on the GR branch solution of the EGB gravity since this branch has a general relativistic limit and it's easy to compare the results with black holes in general relativity. Non-GR branch solution on the other hand behaves quite differently and the results are not so easy to interpret. Since our aim is to compare the effects of Gauss-Bonnet term with general relativity, we have not considered the non-GR branch here.\\
 For dynamical EGB black hole we have shown the surface gravity associated with the trapping horizon is indeed it's temperature through our calculation. There are several different definitions of black hole horizon as well as surface gravity available in the literature for a dynamical spacetime. Among all these, the Hayward-Kodama formalism stands out for it's coherent and consistent description of a dynamical black hole spacetime. Since most of the previous works are done in a $4$-dimensional spacetime, it is not clear a priori how this formalism would work for gravity theories in higher dimensions. And it is reassuring to find that the trapping horizon formalism works nicely for GR branch solution in dynamical EGB gravity and produce the expected results. In \cite{Nozawa}, the first law of black hole dynamics was proved assuming $\kappa_{TH}$ as the temperature of the trapping horizon which we have shown explicitly in this paper. So this is another point in favor of Hayward-Kodama formalism.\\
 The method we have employed to calculate temperature of the dynamical horizon has been used to calculate temperature for the stationary (both non-extremal and extremal)\cite{Chatterjee1} as well as dynamical horizons\cite{Chatterjee2,Chatterjee3}. For dynamical horizons in $4$-dimensions the symmetry of the spacetime plays an important role. The Kodama vector and the Misner-Sharp energy can be defined uniquely only if the spacetime is spherically symmetric and this is essential for our calculations. For dynamical horizons in higher dimensional GB gravity we have chosen the spacetime to be maximally symmetric which ensures the existence of Kodama vector and generalized Misner-Sharp energy. For a different geometry, a Kodama-like vector field is not known yet and its not obvious how our formalism can be extended to these type of spacetimes. \\
 Another important point is that this method calculates the correct temperature for a horizon without resorting to any approximations unlike the Hamilton-Jacobi tunneling formalism \cite{Vanzo}, which depends heavily on WKB approximations. The method is semiclassical in the sense that we use quantum field theory in a classical gravity background but the mode solutions are exact and this method is applicable to both extremal and non-extremal spacetimes \cite{Chatterjee1}.\\
For the $4$-D novel EGB black hole we have calculated the temperature, entropy, heat capacity and free energy associated with the event horizon for the GR branch solution. Though the calculations are straightforward and produce standard result for temperature, the expression for entropy is somewhat surprising. Even with inclusion of higher order curvature term in action, there is no classical example of the existence of the logarithmic term in entropy. It appears only as leading order correction to entropy in quantum theories of gravity. This expression for entropy was found in \cite{Fernandes} for a charged black hole in AdS spaces in $4$-D EGB gravity. Surprisingly both the contribution from the electric charge and the cosmological constant canceled out and the result is same with what we have found for the static spherically symmetric black hole. How such terms are coming into play is not clear at this moment and maybe this unusual theory is an indicator of how these corrections appear in quantum gravity, which remains to be seen.\\
 There are several issues related to the $4$-D EGB gravity theory, the most important being the absence of a valid action and consistent theory in $4$-dimensions\cite{Dadhich,Ai}. As was claimed in the original work \cite{Glavan}, this theory is formulated in $n > 4$ dimensions with a rescaled Gauss-Bonnet coupling constant $\alpha \rightarrow \alpha/(n-4)$  and the $4$-D theory is recovered only by taking limit $\alpha \rightarrow 4$ at the level of equations of motion after variation of action. First of all, this action does not follow from Lovelock gravity which is expected naturally for higher curvature gravity theories. Secondly there are some unnatural consequences which follow from this ad hoc choice of rescaling the Gauss-Bonnet coupling constant. The black hole spacetime in this theory admits two horizons just like a Reissner-Nordstr\"{o}m black hole despite the absence of any electrical charges. What is more peculiar is that the coupling constant $\alpha$, which is a geometrical quantity behaves like an electric charge, a physical quantity, so that the whole causal structure of the spacetime has changed and the central singularity is timelike instead of spacelike. This is not comfortable. On the other hand there is the expression for entropy for this black hole, which indicates though the spacetime looks like Reissner-Nordstr\"{o}m apparently, it is not exactly so. There are some subtleties in this theory which are not fully understood yet. There has been some effort to make a consistent $ n \rightarrow 4$ Einstein-Gauss-Bonnet gravity\cite{Mukhoyama} where it was shown that the regularization is possible only for some broad, but limited, class of metrics and only by breaking (a part of) the diffeomorphism invariance or by adding an extra degree of freedom, in agreement with the Lovelock theorem, a consistent $ n \rightarrow 4$ theory can be constructed which is different from general relativity in $4$-dimension. Whether this unusual theory can genuinely become an alternative to Einstein's general relativity and it's interesting features can provide some insights into quantum theories of gravity remains to be seen.
 \section{Acknowledgements}

 I would like to thank Narayan Banerjee for useful discussions. This work is supported by DST Project Grant No. SR/WOS-A/PM-78/2017 under the DST Women Scientists Scheme A.

\end{document}